\documentclass[10pt]{article}

\usepackage{url}
\usepackage{color}
\usepackage{graphicx}
\usepackage{bm}
\usepackage{amsmath}
\usepackage{amssymb}
\usepackage{cite}

\usepackage{setspace}
\newenvironment{sabstract}{%
\begin{quote} }
{\end{quote}}

\title{\vspace*{-2cm} Shaping nanoparticle fingerprints at the interface of cholesteric droplets} 

\author
{Lisa Tran,$^{1\ast}$ Hye-Na Kim,$^{2}$ Ningwei Li,$^{3}$\\
 Shu Yang,$^{2}$ Kathleen J. Stebe,$^{4}$ Randall D. Kamien,$^{1}$\\
and Martin F. Haase $^{5\ast}$ \\
\\
\\
\small{$^{1}$Department of Physics and Astronomy, University of Pennsylvania,}\\
\small{209 South 33rd Street, Philadelphia, PA 19104, USA}\\
\small{$^{2}$Department of Materials Science and Engineering, University of Pennsylvania,}\\
\small{3231 Walnut Street, Philadelphia, PA 19104, USA}\\
\small{$^{3}$Department of Mechanical and Industrial Engineering, University of Massachusetts,}\\
\small{160 Governors Drive, Amherst, MA 01003, USA}\\
\small{$^{4}$Department of Chemical and Biomolecular Engineering, University of Pennsylvania,}\\
\small{220 South 33rd Street, Philadelphia, PA 19104, USA}\\
\small{$^{5}$Department of Chemical Engineering, Rowan University,}\\
\small{600 North Campus Drive, Glassboro, NJ 08028, USA}\\
\\
\normalsize{$^\ast$To whom correspondence should be addressed;}\\
\normalsize{E-mail: ltran@sas.upenn.edu, haasem@rowan.edu.}
}

\date{}

\begin{document} 



\maketitle 


\begin{sabstract}
The ordering of nanoparticles into predetermined configurations is of importance to the design of advanced technologies. In this work, we moderate the surface anchoring against the bulk elasticity of liquid crystals to dynamically shape nanoparticle assemblies at a fluid interface. By tuning the degree of nanoparticle hydrophobicity with surfactants that alter the molecular anchoring of liquid crystals, we pattern nanoparticles at the interface of cholesteric liquid crystal emulsions. Adjusting the particle hydrophobicity more finely further modifies the rigidity of assemblies. We establish that patterns are tunable by varying both surfactant and chiral dopant concentrations. Since particle assembly occurs at the interface with the desired structures exposed to the surrounding phase, we demonstrate that particles can be readily crosslinked and manipulated, forming structures that retain their shape under external perturbations. This study establishes the templating of nanomaterials into reconfigurable arrangements. Interfacial assembly is tempered by elastic patterns that arise from the geometric frustration of confined cholesterics. This work serves as a basis for creating materials with chemical heterogeneity and with linear, periodic structures, essential for optical and energy applications.  
\end{sabstract}

\newpage


\section*{INTRODUCTION}

\indent \indent The ability to organize nanoparticles into designed arrangements is of interest for a wide range of material applications, including nanomedicine, energy harvesting, catalysis and optical devices \cite{app-nanomed,app-shuyang-goldring}. Recently, structured assemblies of nanoparticles have been achieved within the bulk of polymer matrices \cite{np-polym}, block copolymers \cite{np-bcp}, and liquid crystals \cite{np-lc-lacaze-1,np-lc-lacaze-2,np-lc-mitov,np-lc-hegmann,np-lc-hirst,mitov-1,mitov-2}. However, traditional polymers and block copolymers are not easily reconfigurable with external stimuli, limiting their use in applications that require dynamic responses. Furthermore, although we have demonstrated that liquid crystals do reconfigure with changes in temperature and surfactant concentration \cite{lt-prx}, the assemblies achieved within liquid crystals thus far have nanoparticles dispersed and embedded within the bulk. This limits chemical access to the resultant nanoassemblies, which is essential for their production and use in sensing and optical applications \cite{app-shuyang-goldring, patch-review, tll-waltz-patch, abblipid, abb-patch, lagerw-poly}.

Our previous work has shown that cholesteric liquid crystals can alter their arrangements with changing boundary curvature, developing a fingerprint texture that often has twisted, double spiraled, focal conic domains, in which the texture is comprised of alternating regions of liquid crystal anchoring to a surface \cite{lt-prx}. Further, in order for a cholesteric to produce focal conic domains, it must deform its interface into a hilly topography, with each hill accommodating a double spiral, generating nontrivial interfacial curvature \cite{lt-prx, pieranski-1, pieranski-2, boulig-spherulites}. These phenomena establish an interplay between bulk elasticity and surface tension. It remains unexplored how such interplay shapes nanoparticle assemblies, as nanoparticles can kinetically arrest and aggregate, introducing rigidity to the interface. Can surface active nanoparticles follow liquid crystal patterning? What is the role of nanoparticle surface chemistry in the response of assemblies to the elasticity of liquid crystals? 

In this article, we demonstrate the structuring of surfactants and nanomaterials at the \textit{interface} of cholesteric liquid crystals. We realize interfacial nanoparticle assemblies that are shaped by the elastic field of liquid crystals, a phenomenon previously predicted by simulations \cite{jjdp-np-int-1, jjdp-np-int-2} but is now accomplished experimentally. With the use of surfactant-modified nanoparticles that attach to the water-liquid crystal interface, we create patterned nanoparticle-decorated emulsions at high densities and with crosslinkable, assembled arrangements that are dynamically controllable through the underlying elastic field. This approach is fundamentally different from the assembly of nanomaterials within liquid crystal defects, as it relies not on defects but instead on how the bulk responds to the molecular orientation of liquid crystals to their confining boundaries. This method further exploits the intrinsic ability of cholesteric liquid crystals to form complex, but ordered surface patterns. Moreover, our findings elucidate a previously unexplored regime in particle stabilized emulsion systems: the interfacial adsorption energy of the nanoparticles that we utilize are such that their aggregation is \textit{moderated} by the liquid crystal elastic energy, revealing that their adsorption behavior follows kinetics reminiscent of nucleation and growth.

\section*{RESULTS}

\subsection*{Patterned segregation of lipids and particles at the cholesteric-water interface}

\begin{figure*}[ht!]
\centerline{\includegraphics[width=1\textwidth]{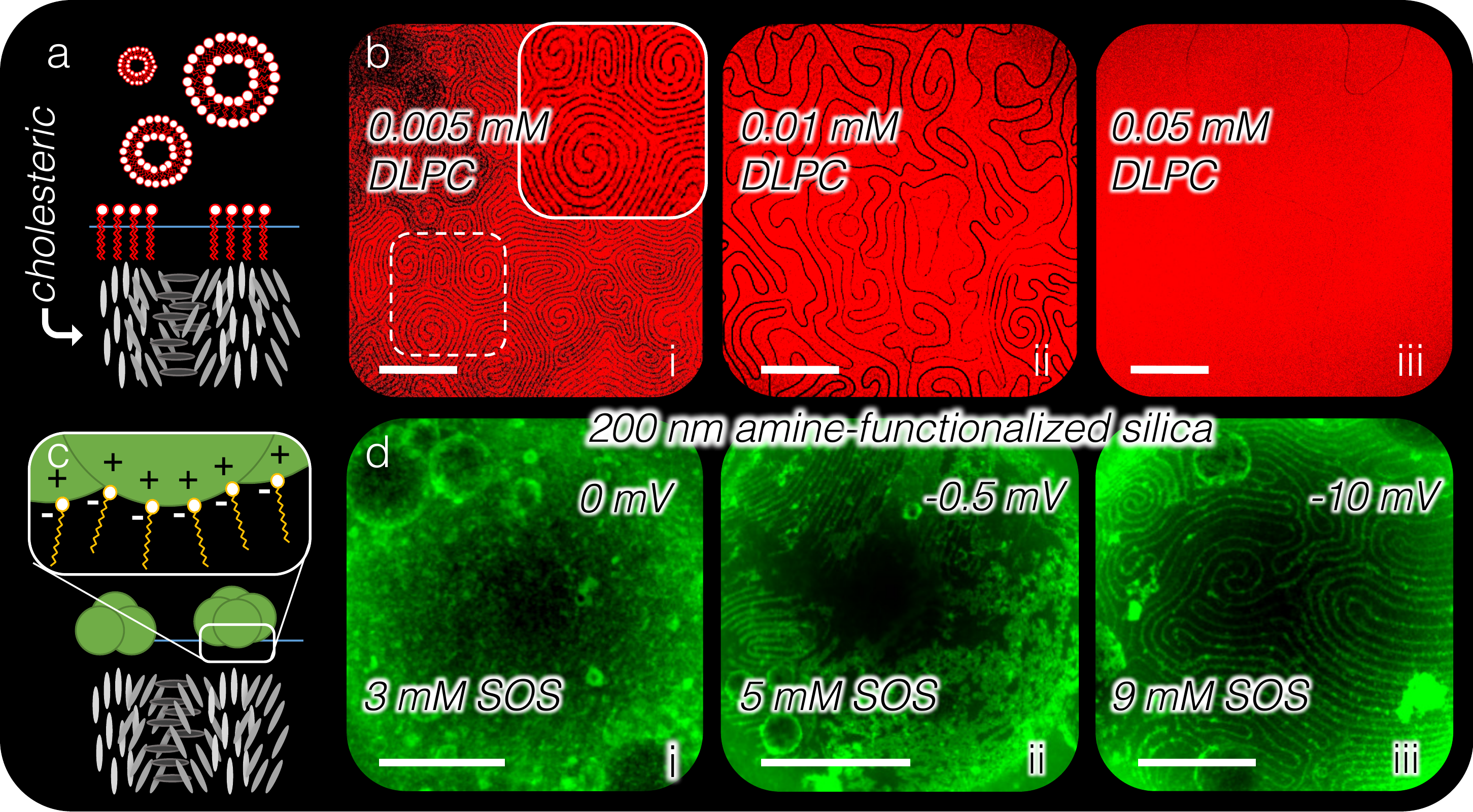}}
\caption{\label{Intro} Lipid and nanoparticle segregation at the cholesteric liquid crystal-water interface. Scale bars are 50 $\mu$m. a) Cholesterics (gray) must twist \textit{along} the surface to have as much homeotropic anchoring as possible from the presence of lipids in the surrounding water phase. The hydrophobic tails of the lipids prefer the liquid crystal phase, causing liquid crystal molecules to lie parallel to the tail and thus perpendicular to the interface. Twist regions with molecules tangent to the interface exclude traditional, molecular surfactants, such as lipids (red). b) Confocal data of lipids (TR-DLPC, red) demonstrates lipid segregation at the cholesteric-water interface. As lipid concentration increases (i-iii), surface stripes become wider and more disordered (ii) until twist regions are forced away from the surface as a result of the lipids saturating the interface (iii). c) Surfactant-decorated nanoparticles, made surface-active from the electrostatic grafting of molecular surfactants to the nanoparticle surface, are also found to align with molecules perpendicular to the interface, forming patterned assemblies. d) Projections of confocal z-stacks of nanoparticles (green) on cholesteric droplets demonstrates how surfactant-modified nanoparticles can follow the underlying cholesteric patterning. Electrostatic surface functionalization allows for flexible surface chemistry. Silica nanoparticles that are amine-functionalized and thus positive in surface charge can be modified by negatively charged SOS to become surface active. Increasing the SOS concentration to obtain nanoparticles with sufficiently negative zeta potentials are needed for the particles to segregate into stripes (iii). Numbers on the upper right corners of micrographs are system zeta potential measurements.}
\end{figure*}

\indent \indent A homogeneous mixture of 5CB (4-cyano-4'-pentylbiphenyl) doped with CB15 ((S)-4-cyano-4- (2-methylbutyl)biphenyl) is emulsified to form droplets in an aqueous phase, stabilized by surfactants that induce homeotropic anchoring. Homeotropic anchoring is achieved through the use of hydrocarbon surfactants with hydrophilic heads and hydrophobic tails. The hydrophobic tails intercalate between liquid crystal molecules, causing them to align parallel to the tail, perpendicular to the interface (Fig.~\ref{Intro}a) \cite{lt-prx, abblipid, abbsurf}. A cholesteric breaks both orientational and translational symmetries, as its molecules have an energetic tendency to not only align with one another, but also to have a bulk twist, stacking molecules in a helical fashion and imparting a periodic phase into the material. When a cholesteric is bounded by a surface that induces homeotropic anchoring, there is no way for the twisting molecules to orient near the boundary without frustrating either the anchoring energy or the twist energy. At moderate homeotropic anchoring strengths, alternating regions of homeotropic and planar anchoring occur at the surface due to periodic violation of the homeotropic boundary condition by the twist elastic energy, as depicted in Fig.~\ref{Intro}a \cite{lt-prx}. Stripes of planar anchoring are created at the surface by the cholesteric twist. We utilized the competition between the cholesteric's energetic preference to twist and a homeotropic boundary condition that aligns molecules perpendicular to the boundary to generate surface stripes.

From the use of the fluorescently labeled lipid TR-DLPC (Materials and Methods), we find that not only do lipids induce homeotropic anchoring of the cholesteric, but the cholesteric subsequently \textit{segregates} the lipids at the interface, excluding them from twist regions incompatible with the anchoring (Fig.~\ref{Intro}b). This generates a system with an inhomogeneous, dynamic boundary condition. At 0.005 mM TR-DLPC, thin stripes form double spiraled focal conic domains, visible under confocal microscopy from the creation of lipid-depleted stripes along twist regions. As the TR-DLPC concentration is increased to 0.01 mM, more lipids adsorb to the interface. The thickness of the lipid-filled stripes increases, disrupting the stripe pattern as the twist energy is further frustrated by the increasing homeotropic anchoring energy. The surface pattern disappears entirely when the interface is fully saturated with lipids --- the homeotropic anchoring is so strong that it prevents the cholesteric from twisting at the surface.

To pattern nanoparticles at the cholesteric interface, we functionalize them with surfactants, (Fig.~\ref{Intro}c) according to the literature \cite{surf-np-1, binks, mfh-bijel}. To this end, ionic surfactants are selected to electrostatically adsorb to the surfaces of positively or negatively charged particles, respectively. This \textit{in-situ} modification of altering nanoparticle hydrophobicity facilitates the use of surfactants that have been proven to change liquid crystal anchoring at the liquid crystal-water interface \cite{abbsurf}. The surfactants used in these experiments, either anionic sodium alkyl sulfate (SA$_n$S) or cationic trimethyl alkylammonium bromide (A$_n$TAB), where $n$ indicates the alkyl chain length, have each been shown to induce homeotropic anchoring for 5CB \cite{lt-prx, abbsurf}. Moreover, this method of surface modification provides flexibility in material composition of the nanoparticles. Two types of fluorescent nanoparticles are used to study the effect of nanoparticle size and surface chemistry: positively charged amine-functionalized, 200-nm particles and negatively charged, bare, 30-nm silica nanoparticles. The former are \textit{in-situ} surface modified with SA$_n$S and the latter with A$_n$TAB, respectively. Similar to the lipids in Fig.~\ref{Intro}b, nanoparticles are able to segregate into stripes that have homeotropic anchoring, but only under specific solution conditions, as depicted in Fig.~\ref{Intro}c and as demonstrated in the confocal micrographs of Fig.~\ref{Intro}d. Nanoparticle concentration, surfactant concentration, nanoparticle size, surfactant tail length, and solution pH are all found to influence the behavior of nanoparticles at the cholesteric-water interface.

The degree of surfactant coverage on nanoparticles determines the particle segregation behavior on cholesteric droplets. Confocal micrographs depicting the behavior of 200-nm, fluorescent, silica nanoparticles on cholesteric droplets are given in Fig.~\ref{Intro}d and Fig.~\ref{S-VaryS8}. The droplets are emulsified by hand shaking in a nanoparticle dispersion of $10^{-3}$ wt-\% with varying SOS (sodium octyl sulfate) concentrations. The degree of surfactant coverage on the nanoparticles is determined by zeta potential measurements of selected samples (Fig.~\ref{Intro}d and Fig.~\ref{S-VaryTail}b) \cite{zeta-text}. The amine-functionalized particles alone have a zeta potential of 36 mV, confirming a highly positive surface charge that provides enough electrostatic repulsion for colloidal stability --- \textit{i.e.} the nanoparticles are well-dispersed and hydrophilic. A solution of these nanoparticles with 3 mM SOS results in a zeta potential of 0 mV, indicating saturated monolayer adsorption of SOS on the particles. At these low SOS concentrations (8 mM and less), the nanoparticles become hydrophobic and strongly aggregate at the cholesteric interface and in the solution (Fig.~\ref{Intro}d, i-ii). Increasing the SOS concentration slowly decreases the zeta potential. This indicates the onset of SOS double layer formation on the particles with increasing SOS concentration, with negatively charged sulfonate groups oriented towards the aqueous solution. Indeed, it is only around and slightly above 10 mM that the nanoparticles are better dispersed and well ordered within the homeotropic stripes of the cholesteric droplet (Fig.~\ref{Intro}d-iii; Fig.~\ref{S-VaryS8}, green). The degree of hydrophobicity of the nanoparticles must decrease for them to aggregate less, with their mutual repulsion facilitated by the slight surfactant double layer. However, decreasing the nanoparticle hydrophobicity further by increasing the SOS concentration to 25 mM results in fewer nanoparticles at the droplet interface, suggesting that the double layer has progressed such that the nanoparticle surface charge is dominated by hydrophilic head groups, causing the nanoparticles to become hydrophilic again. The surfactant concentration must be moderated to enable both particle dispersion and particle adsorption onto the liquid crystal-water interface.

\subsection*{Effect of particle hydrophobicity, particle size, and pH of solution}

\begin{figure*}[ht!]
\centerline{\includegraphics[width=1\textwidth]{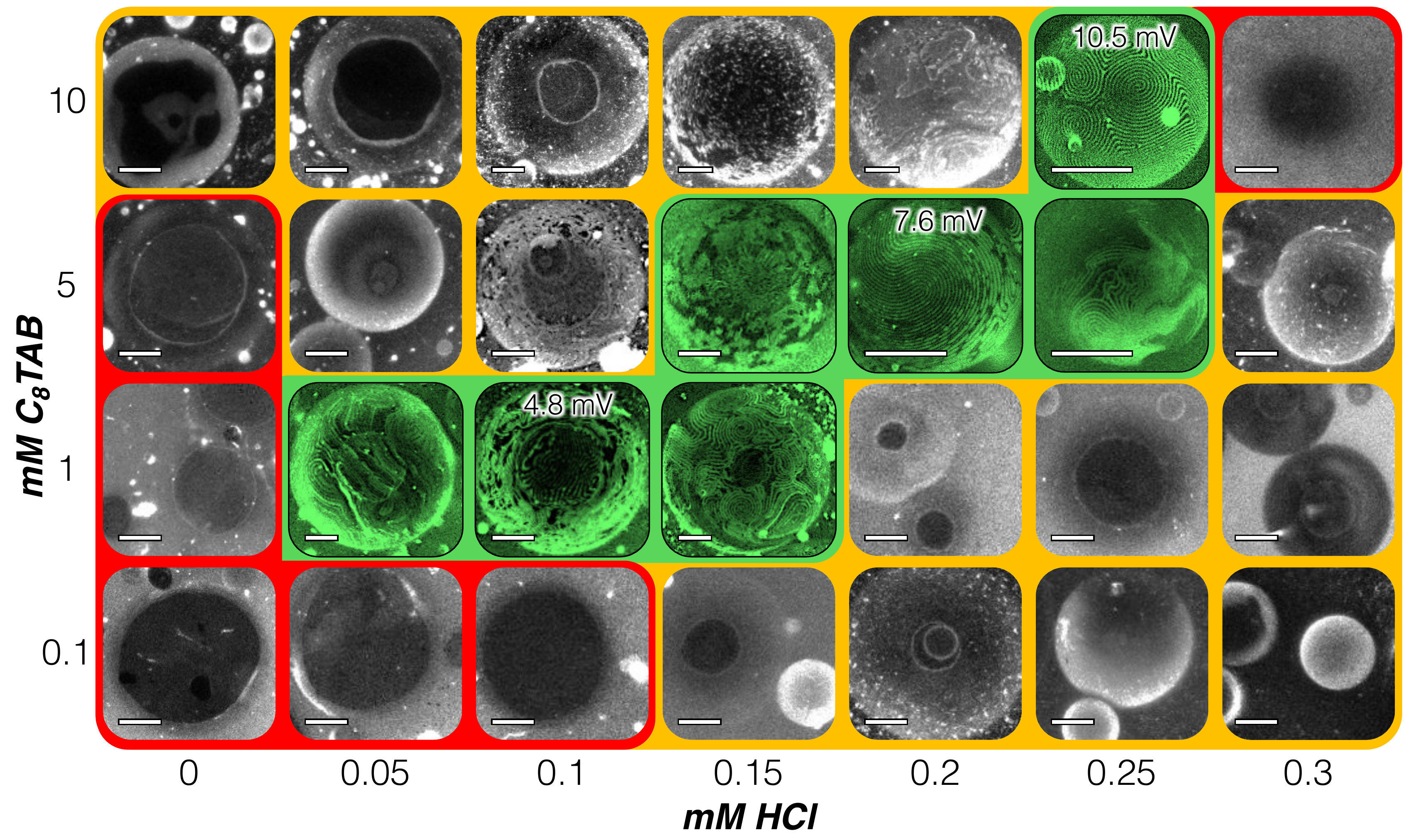}}
\caption{\label{PhaseDiag} State diagram for nanoparticle stripe segregation with varying concentrations of HCl and C$_{8}$TAB. Scale bars are 50 $\mu$m. Segregation of surface modified nanoparticles takes place in a narrow pH and surfactant concentration range. Silica nanoparticles (30 nm) are \textit{in-situ}, surface functionalized with C$_{8}$TAB. Droplets of the cholesteric 5CB with 3 wt-\% CB15 are formed by simple vial shaking. Varying the concentration of C$_{8}$TAB and HCl shows three regions: Red regions indicate that particles do not attach to the cholesteric interface. Yellow regions signify unordered, interfacial assembly of nanoparticles. Green regions denote conditions where particles are surface active \textit{and} have cholesteric ordering, forming stripes. Zeta potentials for selected systems are given at the top of micrographs. The bare silica nanoparticles have a slightly positive zeta potential that becomes more positive with increasing C$_{8}$TAB concentration, implying the presence of a slight C$_{8}$TAB double layer at their surface that further develops with increasing concentration.} 
\end{figure*}

For the uniform segregation of surfactant-modified nanoparticles into homeotropic stripes, we identify three additional criteria:

(i) Since the degree of hydrophobicity is key to nanoparticle dispersion and wetting behavior, it follows that the \textit{length of the surfactant tail} also affects the nanoparticle behavior, exhibited in Fig.~\ref{S-VaryTail} \cite{pickering-akartuna}. Shorter surfactant tail lengths offer greater colloidal stability, enabling nanoparticles to be dispersed enough for them to adsorb to the cholesteric interface with minimal aggregation. This in turn enables particles to better follow homeotropic stripes, and to fully deplete from planar, twist regions. Because of this, we mainly use surfactants with a tail length of eight carbons for nanoparticle surface modification.

(ii) The nanoparticle \textit{size} also affects their aggregation behavior. At a fixed surfactant and nanoparticle concentration, the 200-nm particle system has a greater degree of particle aggregation compared to those with 30-nm particles. Working with smaller nanoparticles is thus more advantageous as the particle aggregation is better managed by larger particle surface areas and greater thermal activity. To this end, we use bare, untreated, 30-nm silica particles with C$_8$TAB.




(iii) The solution \textit{pH} further influences nanoparticle behavior, as it affects the amount of surfactants adsorbed onto the nanoparticle surface \cite{mfh-pickering, zeta-text}. Zeta potential measurements with varying pH in the presence of either C$_8$TAB for bare silica particles or SOS for amine-functionalized particles are given in Fig.~\ref{S-Zeta}.



Considering the above, we explore the interfacial behavior for 30-nm, bare silica nanoparticles with varying C$_8$TAB and HCl concentrations in Fig.~\ref{PhaseDiag}. In this state diagram, red indicates no interfacial wetting of the nanoparticles, yellow indicates interfacial attachment but with aggregation, and green indicates interfacial attachment with particle segregation into cholesteric patterns. 


Similar to the 200-nm, amine-functionalized system in Fig.~\ref{Intro}d, segregation of nanoparticles into stripes occurs only within a narrow regime, where the hydrophobicity of the particles is not so great that the particles aggregate with one another both in solution and at the interface, but also not so small that they do not adsorb. At low surfactant concentration (from 0.1-1 mM C$_8$TAB), the more basic the solution is, the more negatively charged silanol groups there are on the nanoparticle surface. At these conditions, no significant particle deposition on the droplets can be observed (red region). The resulting droplets wet the surface of the microscope slide, creating black regions in the middle of droplets in confocal micrographs. Decreasing the pH by adding HCl increases the nanoparticle hydrophobicity, bringing the system from non-attachment (red), to stripe segregation (green), to randomly organized nanoparticle aggregates (yellow), suggesting increasing hydrophobicity with decreasing pH. However, at a high surfactant concentration (from 5-10 mM C$_8$TAB), the system starts off already hydrophobic at high pH (yellow) and crosses over to stripe segregation (green) and then non-attachment (red) with additions of HCl. This is consistent with the hypothesis that, at high surfactant concentration, a slight surfactant double layer is needed at the particle surface for successful segregation within stripes. Again, too complete of a surfactant double layer will make the particles too hydrophilic, resulting in no aggregation and no interfacial attachment, as seen with 10 mM C$_8$TAB and 0.3 mM HCl (red). 

By moderating C$_8$TAB and HCl concentrations to adjust particle hydrophobicity, nanoparticles can follow the cholesteric pattern, with the best particle ordering seen at 10 mM C$_8$TAB and 0.25 mM HCl. The increasing zeta potential of green regions with increasing C$_8$TAB concentration provides further evidence for the presence of a surfactant double layer on the nanoparticle surface. Indeed, before nanoparticle attachment, stripes are evident on droplets under the surfactant and acid conditions of all systems within green regions of the state diagram. This implies that, in addition to forming a slight double layer on particle surfaces, surfactants alter the cholesteric anchoring before particle adsorption to the interface. Interestingly, the stripes formed from 1 to 5 mM C$_8$TAB with HCl concentrations ranging from 0 to 0.2 mM, have nanoparticle stripes that do not perfectly follow the cholesteric ordering. Instead, the assemblies appear rigid and have regions with aggregated particles. Only for 5 mM and 10 mM C$_8$TAB, with 0.25 mM HCl, do the nanoparticle assemblies better follow the cholesteric ordering, matching the lipid results of Fig.~\ref{Intro}b. Exact acid concentrations may vary depending on the age of solutions, as these systems are pH sensitive and become more acidic with time due to the absorption of carbon dioxide. 

\subsection*{Varying the width of nanoparticle-decorated stripes}

\begin{figure*}[ht!]
\centerline{\includegraphics[width=1\textwidth]{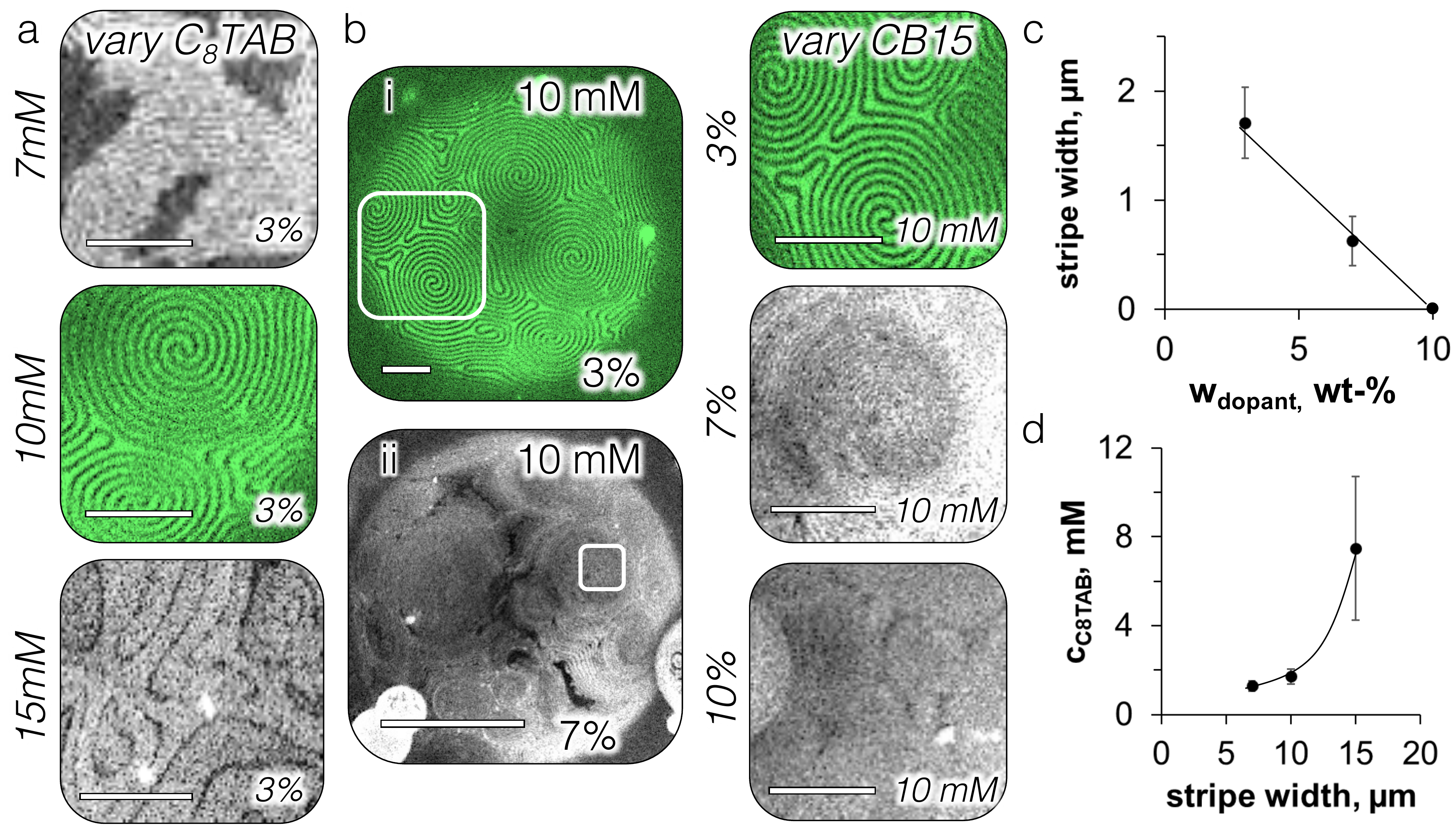}}
\caption{\label{StripeW} Controlling the thickness of nanoparticle-filled stripes with CB15 or C$_{8}$TAB concentrations. Scale bars are 25 $\mu$m. a,d) Similar to the lipid results in Fig.~\ref{Intro}b, the nanoparticle-filled stripe width can also be controlled with the C$_{8}$TAB concentration. b,c) The stripe width can be tuned by adjusting the concentration of the chiral dopant, CB15. By increasing the chiral dopant from 3\% to 7 wt-\% at a fixed concentration of 10 mM C$_{8}$TAB, the stripe width decreases from $\sim$1.7 $\mu$m to $\sim$600 nm, corresponding to a decrease in the cholesteric pitch (a). Increasing the dopant concentration to 10 wt-\%, the cholesteric pitch decreases, with a projected surface stripe width on the order of $\sim$100 nm. However, no stripe segregation is evident from confocal data. Instead, nanoparticles organize into circular domains dictated by cholesteric double spiral domains (bottom right of b). Lines in c) and d) are drawn to guide the eye.} 
\end{figure*}

To better understand the nanoparticle behavior at these conditions, we adjust the nanoparticle stripe width by varying the amount of C$_8$TAB at fixed dopant concentration (Fig.~\ref{StripeW}a,d). Similar to the lipids system in Fig.~\ref{Intro}b, increasing the C$_8$TAB concentration from 10 mM to 15 mM in the solution can increase the stripe thickness, but the stripes also become more disordered as the homeotropic anchoring energy further frustrates and distorts the cholesteric twist, producing disordered stripe patterns (Fig.~\ref{StripeW}a, bottom). Since the lipids in Fig.~\ref{Intro}b are still fluid at the interface, the similar behavior between the lipid and nanoparticle systems suggests that, at these conditions, the nanoparticles may also be fluid. This is unusual for nanoparticle-decorated emulsions, as Pickering emulsions are typically stabilized by rigid interfacial particle films with kinetically arrested particles at the interface \cite{surf-np-1, binks, mfh-bijel}. However, fluid interfacial behavior of nanoparticles has been reported previously for gold nanoparticles capped with (1-mercaptoundec-11-yl)tetra(ethylene glycol) \cite{npmobil-stebe} and for silica nanoparticles functionalized with A$_{n}$TAB with varying surfactant and salt concentrations \cite{npmobil-walker,npmobil-binks}. Our solution conditions have fine-tuned the nanoparticle hydrophobicity by adjusting surfactant concentration and pH such that the particles have both interfacial activity and colloidal stability, facilitating their mobility at the cholesteric interface.

Maintaining 10 mM C$_8$TAB and 0.25 mM HCl, the nanoparticle stripe width can also be adjusted by tuning the pitch of the cholesteric twist. This can be accomplished by modifying the chiral dopant (CB15) concentration within 5CB, as demonstrated in the confocal micrographs in Fig.~\ref{StripeW}b,c. All measurements discussed thus far have been performed with 3 wt-\% CB15, giving a cholesteric pitch of $\sim$5 $\mu$m, measured with the Grandjean-Cano wedge cell \cite{Cwedge,GJWedge}. Surface stripe widths are equal to half of the cholesteric pitch with unfrustrated twisting at the boundary, giving stripe widths of $\sim$2.5 $\mu$m for 3\% CB15 \cite{lt-prx}. Increasing the dopant concentration to 7\% decreases the size of the pitch to slightly above 1 $\mu$m, giving a stripe width of around 600 nm (Fig.~\ref{StripeW}c). However, increasing the chiral dopant concentration further to 10\%, with a cholesteric pitch of $\sim$600 nm and an expected surface stripe width of $\sim$300 nm, results in no stripes visible under the confocal microscope (Fig.~\ref{StripeW}b, bottom right). Instead, the cholesteric ordering appears as circular domains that map out the locations of cholesteric double spirals. It is intriguing that the fluorescent intensity is lower in the regions between circular domains, indicating fewer nanoparticles located at these areas, as seen also for 7\% CB15 in between double spirals (Fig.~\ref{StripeW}b-ii). It is also possible that the confocal resolution is not high enough to resolve the stripe width for the 10\% CB15 system. It would be interesting to view the surface of these droplets with AFM or SEM, but polymerizing or otherwise solidifying the liquid crystal would be necessary to perform such measurements \cite{lagerw-poly}. More detailed studies are necessary. Our method of templating nanoparticle interfacial assemblies with liquid crystals still provides a plausible method to create periodic nanoparticle structures on the range of hundred of nanometers, a length scale that is often difficult to obtain with other techniques.



\subsection*{Nanoparticle adsorption dynamics}

\begin{figure*}[ht!]
\centerline{\includegraphics[width=1\textwidth]{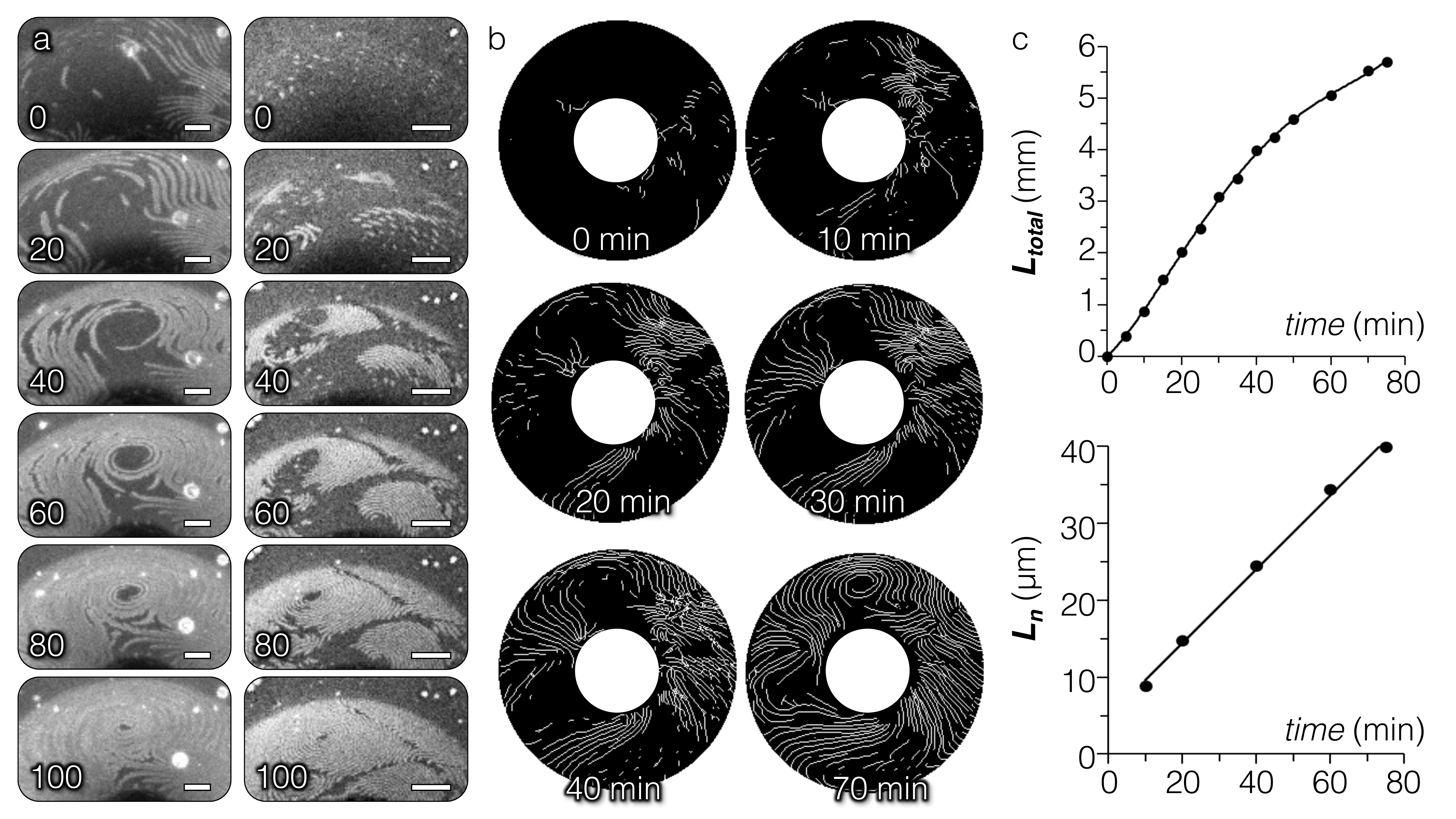}}
\caption{\label{TimeEvo} Time evolution of hydrophobic nanoparticles coating a cholesteric droplet. Confocal data of nanoparticles coating two droplets with differing stripe widths are shown in a), where the left column is a zoom in of the top of the droplet in b). Scale bars are 25 $\mu$m. Preassembled nanoparticle clusters translate and coalesce along stripes that result from the cholesteric ordering of the droplet. The total stripe length growth rate is highest at the beginning of the interfacial attachment process and decreases with time due to the saturation of the surface with nanoparticle-filled stripes. The total stripe length of all nanoparticle stripes on the droplet shown in the left of a) and in b) is plotted against time in c), where the total length is given by $L_{total} = \sum N_{i} \cdot L_{i}$, where $N_{i}$ is the number of stripes with the length $L_{i}$. The total stripe length normalized by the number of stripes $N_{i}$ with the length $L_{i}$ is given by $L_{n} = (\sum N_{i} \cdot L_{i}) / \sum N_{i}$, which, when plotted against time, is shown in d) to have a linear growth. Lines for c) are drawn to guide the eye. Similar dynamics are seen in droplets coated with thinner nanoparticle stripes, shown in the right column of a). Video of this process is in Supplementary Information.} 
\end{figure*}

Visualizing the dynamics of nanoparticle deposition on the liquid crystal droplets provides further insights into their interfacial behavior. Decreasing the HCl concentration slightly, from 0.25 to 0.23 mM, while maintaining particle and C$_8$TAB concentrations of 0.01 wt-\% and 10 mM, respectively, triggers the nanoparticles to slowly aggregate with one another, becoming rigid and forming a crust at the interface after two hours. The process leading to crust formation around the cholesteric droplet yields surprising dynamical behaviors, as shown in Fig.~\ref{TimeEvo}. 

The evolution of stripe lengths in Fig.~\ref{TimeEvo}a is measured in Fig.~\ref{TimeEvo}b and Fig.~\ref{S-TimeSeries}a. The number of stripes, $N_i$, of the length, $L_i$, is plotted for each frame of the image sequence in Fig.~\ref{S-TimeSeries}b. At the beginning of the process, there is a large number of short nanoparticle rafts that act as seeds for stripe growth, visible in frames 0 to 20 in Fig.~\ref{TimeEvo}a. These rafts then slide along stripes until they snap together, growing the stripes rapidly in the longitudinal stripe direction, with a slower lateral growth rate (Fig.~\ref{TimeEvo}a, frames 20-100). As the nanoparticle coverage of the droplet progresses, the number of stripe seeds decreases (Fig.~\ref{S-TimeSeries}b, left), as the existing stripes continue to grow in length. The total length of the stripes, $L_{total} = \sum N_{i} \cdot L_{i}$, grows with time and only begins to plateau when limited by the surface area of the droplet (Fig.~\ref{TimeEvo}c, top). The number averaged length of the stripes also shows a positive, linear trend with time, further indicating that growth sites form in open space and elongate as more nanoparticles adsorb to the interface (Fig.~\ref{TimeEvo}c, bottom) \cite{kinetic}. The homeotropic stripes are filled in first, after which the stripes continue to expand as the nanoparticles aggregate further to fully cover the cholesteric droplet.

\subsection*{Crosslinking nanoparticles into cholesteric textures}

\begin{figure*}[ht!]
\centerline{\includegraphics[width=1\textwidth]{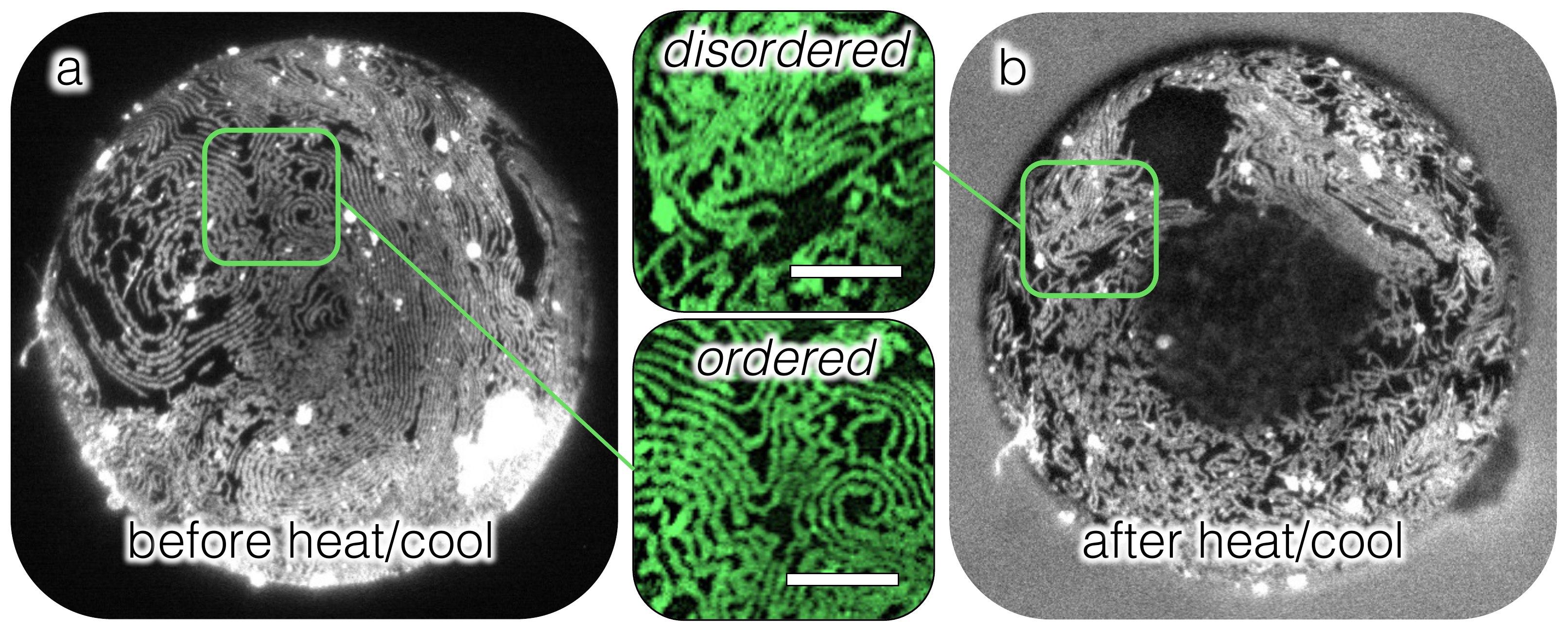}}
\caption{\label{HeatCool} Crosslinking nanoparticle assemblies and destroying cholesteric ordering with temperature. Scale bars are 25 $\mu$m. a) A confocal micrograph shows the crosslinked nanoparticles on a cholesteric droplet. The integrity of the crosslinked nanoparticle stripes is tested by rapidly heating the droplet to and cooling from the isotropic phase to disrupt the cholesteric ordering. b) The confocal micrograph of the droplet in a) after quenching reveals that the nanoparticle assemblies are indeed more disordered, but they still retain their linear shape, confirming their robust structure after crosslinking.} 
\end{figure*}

The role of the cholesteric in structuring nanoparticle assemblies should not greatly affect the dynamics of nanoparticle interfacial attachment, as the general kinetics are likely dictated by the nanoparticles' mutual van der Waal's attraction, electrostatic repulsion, and their interfacial attachment energy. Interfacial coverage of an isotropic oil droplet could also exhibit these nucleation-and-growth-like behaviors. However, the \textit{elastic instabilities} seen in particle film growth on curved surfaces would be greatly affected, as the cholesteric shapes the nanoparticle assemblies to be within stripes, making the growth of particle assemblies one-dimensional instead of the typical two-dimensional surface growth \cite{kinetic}. Investigating how the behavior of assembly growth on a surface is affected when the growth is predominantly one-dimensional would be a compelling future investigation. Past studies of particle assembly growth on surfaces have been done with particles on the many-micron scale \cite{chaikin-part-cryst,colloidcryst, vm-sphere}. The cholesteric acts as a sensor to reveal the particles' interfacial behavior --- \textit{i.e.} whether the particles form structures that are more fluid or more rigid --- facilitating the observation of particle deposition and growth on the nanoscale.

The cholesteric not only senses nanoparticle interfacial behavior, but it also \textit{templates} them. Since the nanoparticle assemblies are in contact with the water phase, they are physically crosslinkable by simple chemical additions to the surrounding solution. After aligning nanoparticle assemblies at 5 mM C$_{8}$TAB and 0.2 mM HCl, we introduce 0.5 mM lanthanum chloride (LaCl$_{3}$) to the solution. LaCl$_{3}$ strongly decreases the electrostatic repulsion between particles and specifically adsorbs onto the silica surface, likely facilitating electrostatic bridging, resulting in silica nanoparticles strongly binding to one another \cite{lacl3-crosslink}.



The integrity of the crosslinked nanoparticle assemblies is tested by heating and cooling the cholesteric droplet to and from the isotropic phase (Fig.~\ref{HeatCool}). This quenches the cholesteric, disrupting the bulk ordering and disturbing the surface pattern formed by the cholesteric. Comparison of a cholesteric droplet before (Fig.~\ref{HeatCool}a) and after (Fig.~\ref{HeatCool}b) this quenching reveals that the cholesteric pattern has been thrown into disarray. However, the nanoparticle assemblies clearly retain their linear structure, demonstrating that the cholesteric can successfully template nanoparticles into crosslinkable wires. Without crosslinking by LaCl$_3$, the nanoparticle assemblies generally do not preserve their linear structure after quenching the cholesteric droplets (Fig.~\ref{S-HeatCool}).

\section*{DISCUSSION}

\indent \indent In summary, the elastic energy of liquid crystals can mitigate and shape the interfacial assembly of surface-active nanoparticles. Here, we demonstrate cholesteric liquid crystals can sense the rigidity of nanoparticle assemblies at the interface via comparison of assembly structures to surface patterns seen with fluid, molecular surfactants. We show that cholesterics can also serve as templates for nanoparticles to be molded into arrangements with tunable length scales down to hundreds of nanometers, by modifying both the surfactant and the chiral dopant concentrations. We establish that such arrangements are also crosslinkable. The properties of assemblies can be further customized since nanoparticle surface modification is accomplished via simple, \textit{in-situ} electrostatic adsorption. Many materials can have equally facile surface modifications, and silica nanoparticles with alternative compositions at their core can also be synthesized. Moreover, for nanoparticles to locate only in regions of favorable anchoring, the particles are required to strike a balance in their hydrophobicity, to be both interfacially active, yet mutually repulsive. A system with such characteristics is uncharted territory in the study of nanoparticle-decorated emulsions. Our findings reveal that the nanoparticles could plausibly remain fluid at the interface and could further have dynamics similar to nucleation and growth kinetics. This work establishes liquid crystal patterned, nanoparticle-decorated emulsions as novel systems that combine interfacial phenomena with elasticity to design structures with potential use in applications ranging from sensors to optical devices.

\newpage

\section*{MATERIALS AND METHODS}
\indent \indent \textit{Materials:} Amine-functionalized, fluorescent dye-coupled, 200-nm silica nanoparticles are synthesized by following the procedure described below. The materials for this procedure, fluorescein isothiocyanate isomer I (FITC), (3-aminopropyl)trimethoxysilane (APTMS, 97\%), (3-aminopropyl) triethoxysilane (APTES, 99\%), an aqueous solution of sodium hydroxide (NaOH, 0.1M), and tetraethylorthosilicate (TEOS, 98\%), are purchased from Sigma-Aldrich. Ammonium hydroxide (NH$_4$OH, 30\%) is purchased from Fisher Scientific. Bare, fluorescent-core, 30-nm silica nanoparticles are purchased from Creative Diagnostics and are suspended in an aqueous solution at pH 6. Sodium alkyl sulfate (SA$_n$S) and trimethyloctylammonium bromide are also obtained from Sigma-Aldrich. The lipids 1,2-dilauroyl-sn-glycero-3- phosphocholin (DLPC), labeled with 1 mol\% Texas Red 1,2-dihexadecanoyl-sn-glycero-3- phosphoethanolamine, triethylammonium salt (TR-DHPE), are obtained from Avanti Polar Lipids. Hydrochloric acid (HCl), used for adjusting solution pH, is also obtained from Fisher Scientific. For the cholesteric liquid crystal, we use 5CB (4-cyano-4'-pentylbiphenyl, Kingston Chemicals Limited) doped with CB15 ((S)-4-cyano-4-(2-methylbutyl)biphenyl, EMD Performance Materials and Sython Chemicals) for a right-handed cholesteric pitch. Glass surfaces brought into contact with surfactant-functionalized nanoparticles are rinsed with 2 wt-\% poly(diallyldimethylammonium chloride) (PDADMAC, $MW = 200-300$k g/mol, Sigma-Aldrich) in a solution of 0.5 M sodium chloride (NaCl, Fisher Scientific). This treats the glass to minimize emulsion droplets adhering to it. 

\textit{Optical Characterization:} The main confocal microscopy system used in experiments is an Olympus IX73 microscope coupled with a Thorlabs confocal microscopy upgrade. A blue laser (488nm) is used for fluorescent dye excitation. The fluorescent light is collected through a 25 $\mu$m pinhole and passed through optical filters, transparent for wavelengths from 505 - 550 nm for FITC labelled nanoparticles on the detector. The software ThorImage 3.1 is used for image acquisition. Confocal z-stacks are obtained by a motorized focus control, with z-stack step sizes of 1 $\mu$m. ImageJ is used for the 2D projections of the confocal z-stacks. For lipid measurements, confocal micrographs are obtained using an upright Leica TCS SP5 microscope. DLPC labeled with 1 mol \% TR-DHPE is used to determine the surfactant location on the CLC-water interface. A scanning laser wavelength of 543 nm is used to excite TR-DHPE. 

\textit{Stripe Growth Image Analysis:} Confocal z-stack projections are thresholded and binarized in ImageJ. Manual adjustments to separate pixels from adjacent stripes are done when necessary. To measure the total stripe length, stripes are skeletonized (reduced to 1 pixel width), and the number of pixels is measured and converted to micrometers. To measure the length of individual stripes, the image width and height of the skeletonized images is quadrupled to increase the width of the stripes again. The ImageJ feature ``Analyze Particles" is employed to measure the area of each individual stripe, which is used to calculate the length.

\subsection*{Fluorescent Lipid Patterning on a Thin Cholesteric Film} 

\indent \indent With 1,2-dilauroyl-sn-glycero-3- phosphocholin (DLPC, Avanti Polar Lipids), labeled with 1 mol\% Texas Red 1,2-dihexadecanoyl-sn-glycero-3- phosphoethanolamine, triethylammonium salt (TR-DHPE), as the surfactant, the water phase is a tris-buffered saline solutions (10 mM Tris, Fisher Scientific; 0.1M NaCl, adjusted to pH 8.9 with hydrochloric acid (HCl), Fisher Scientific) with a dispersion of vesicles 50 nm in diameter, following the procedure of previous work \cite{abblipid}. A similar procedure from this work is also used for creating a thin cholesteric film in a copper TEM grid on a cover glass treated with octadecyltrimethoxysilane (OTS, Sigma-Aldrich).

\subsection*{Liquid Crystal Pickering Emulsion Preparation} 

\indent \indent All glass surfaces in contact with the emulsions, from the vials, to the slides, to the transfer pipettes (Fisher Scientific) are all treated with PDADMAC before use, as described under \textit{Materials}. Samples are always made with fresh surfactant stock solutions to minimize the effects of hydrolysis and are measured to have a pH range of 7-7.5. When titrating components into the solution, nanoparticles are diluted first, then the pH is adjusted by additions of either HCl or NaOH. This is vortexed before the surfactant is added. The nanoparticle solution is then bath sonicated for 30 minutes before the cholesteric liquid crystal is introduced. Approximately 10 $\mu$L of the liquid crystal is pipetted to 1 mL of the nanoparticle solution. Liquid crystal in water emulsions are then created by simple shaking. Samples are viewed under the confocal microscope in a covered, hydrated container to minimize evaporation.

\subsection*{Preparation of dye-coupled, 200-nm Silica Nanoparticles} 

\indent \indent This procedure is done following the steps outlined in \cite{dye-couple-si-np}. Briefly, fluorescein isothiocyanate isomer I (FITC, Sigma-Aldrich) molecules are covalently bonded with (3-aminopropyl)trimethoxysilane (APTMS, Sigma-Aldrich). First, 0.0015 grams of FITC is dissolved in 2 mL of ethanol and mixed with 0.237 mL of APTMS for 12 hours under stirring with a Teflon-coated magnetic stir bar. Meanwhile, silica nanoparticles with a diameter of 200 nm (dispersed in deionized water), purchased from General Engineering \& Research (San Diego, CA), are re-dispersed in ethanol in a sonication bath for 1 hour. Then, 32.5 mL of an ethanol suspension containing 1 wt-\% of silica particles is mixed with 2.755 mL of ammonia for 10 minutes. Then, 0.689 mL of 0.1 M NaOH aqueous solution is poured into the bath to activate the silanol groups on the particle surface.

To couple the dye to the particles, 208 $\mu$L of as-prepared, FITC-APTMS solution is added to the silica suspension. After 5 minutes of thorough mixing, 40 $\mu$L of tetraethylorthosilicate (TEOS, Sigma-Aldrich) is added drop-wised, and the mixture is reacted for 22 hours under stirring. To remove unreacted dye molecules, the resultant dye-coupled particles are washed with ethanol three times by centrifuging and replacing the supernatant with fresh ethanol.

\subsection*{Amine-functionalization of Dye-Coupled Silica} 

\indent \indent The procedure is done following the steps outlined in \cite{amine}. Briefly, for the functionalization of silica particles, 0.23 grams of silica pellet (as prepared above) is redispersed in 40 mL of fresh ethanol in a sonication bath. Under stirring with a Teflon coated magnetic bar, 4 mL of ammonia solution (0.727 M in ethanol) is added. 170 $\mu$L of APTES solution (0.011 M in ethanol) is then slowly added drop-wised to the silica dispersion. After 15 hours, the resultant solution is washed with ethanol five times by centrifugation to remove the unreacted APTES molecules.

\subsection*{Lanthanum Chloride Cross-Linking of Silica Nanoparticles within Stripes} 

\indent \indent Nanoparticles are aligned within stripes on the surface of cholesteric droplets in a solution of water with 0.01 wt-\%, 30-nm nanoparticles, 5 mM C$_{8}$TAB, and 0.2 mM HCl (Fig.~\ref{PhaseDiag}). Excess nanoparticles are removed by gently replacing the supernatant with a solution of water with only 5 mM C$_{8}$TAB and 0.2 mM HCl. This is repeated three times. The nanoparticles at the interface are then crosslinked within stripes by replacing the supernatant with an aqueous solution with 5 mM C$_{8}$TAB, 0.2 mM HCl, and 0.5 mM lanthanum chloride (LaCl$_{3}$). After leaving the sample to sit for $\sim$5 minutes, the rinsing procedure is performed again with a similar aqueous solution that excludes the LaCl$_3$.

\setcounter{figure}{0}
\makeatletter 
\renewcommand{\thefigure}{S\@arabic\c@figure}
\makeatother


\newpage

\section*{ACKNOWLEDGEMENTS}

A. Dang, G. Durey, S. Hann, E. Horsley, E. Lacaze, D. Lee, T. Lopez-Leon and Y. Xia are acknowledged for materials and helpful discussions. We thank S.S. Margulies and D.F. Meaney for access to confocal microscopy and G. Gray Lawrence for help with confocal measurements. \textbf{Funding:} This work was supported by National Science Foundation Materials Research Science and Engineering Centers (NSF MRSEC) Grant No. DMR-1720530. L.T. acknowledges support from an American Fellowship grant from the American Association of University Women. M.F.H. is supported by National Science Foundation CAREER Award No. CBET-1751479. R.D.K. was partially supported by a Simons Investigator grant from the Simons Foundation. \textbf{Competing interests:} The authors declare that they have no competing interests. \textbf{Data and materials availability:} All data needed to evaluate the conclusions in the paper are presented in the paper and/or the Supplementary Materials. Additional data related to this paper may be requested from the authors.


\newpage

\bibliography{PatternedPickeringBib}
\bibliographystyle{ieeetr}

\newpage

\setcounter{page}{1}

\section*{SUPPLEMENTARY MATERIALS}

\noindent Video S1: Video of Fig.~\ref{TimeEvo}a, constructed from confocal micrographs, depicting nanoparticles adsorbing onto cholesteric liquid crystal droplets in a solution with 0.01 wt-\% of 30-nm, bare silica particles, 10 mM C$_8$TAB and 0.23 mM HCl. 

\begin{figure*}[ht!]
\centerline{\includegraphics[width=0.5\textwidth]{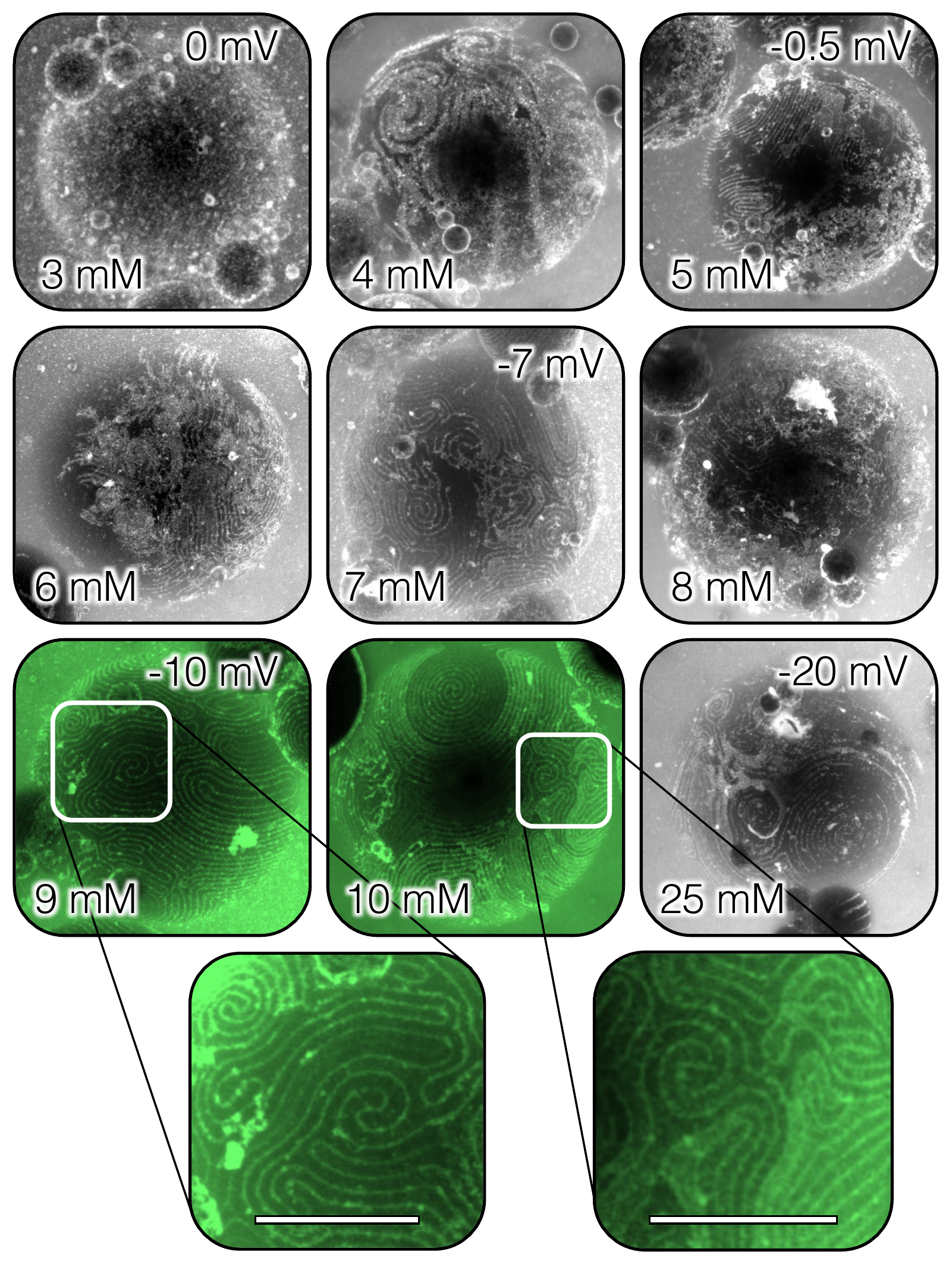}}
\caption{\label{S-VaryS8} Behavior of amine-functionalized silica nanoparticles, modified by varying concentrations of SOS, on cholesteric liquid crystal droplets in water (pH 7). Scale bars are 50 $\mu$m. 200-nm silica nanoparticles are surface-modified by an aminosilane to give the particles an overall positive charge. At the native pH 7 of the particle solution, different concentrations of sodium octyl sulfate (SOS) are added and sonicated before introducing cholesteric liquid crystal. Simple shaking creates droplets stabilized by SOS and the nanoparticles. Confocal micrographs reveal how the nanoparticle behavior changes with increasing SOS concentration. At low SOS concentrations, the nanoparticles aggregates and form crusts on the cholesteric droplets. However, as the SOS concentration increases, the aggregation decreases and the influence of the cholesteric becomes more apparent --- the nanoparticles form assemblies at the interface that follow the cholesteric structures. The greatest degree of nanoparticle ordering is seen at 9 and 10 mM SOS. Zeta potentials of select samples are given at the upper right of micrographs. The slightly negative values suggest that the amine-functionalized nanoparticles have enough of a SOS double layer at its surface to both wet the liquid crystal and to remain repulsive enough to segregate into more energetically favorable, homeotropic stripes.}
\end{figure*}

\begin{figure*}[ht!]
\centerline{\includegraphics[width=0.9\textwidth]{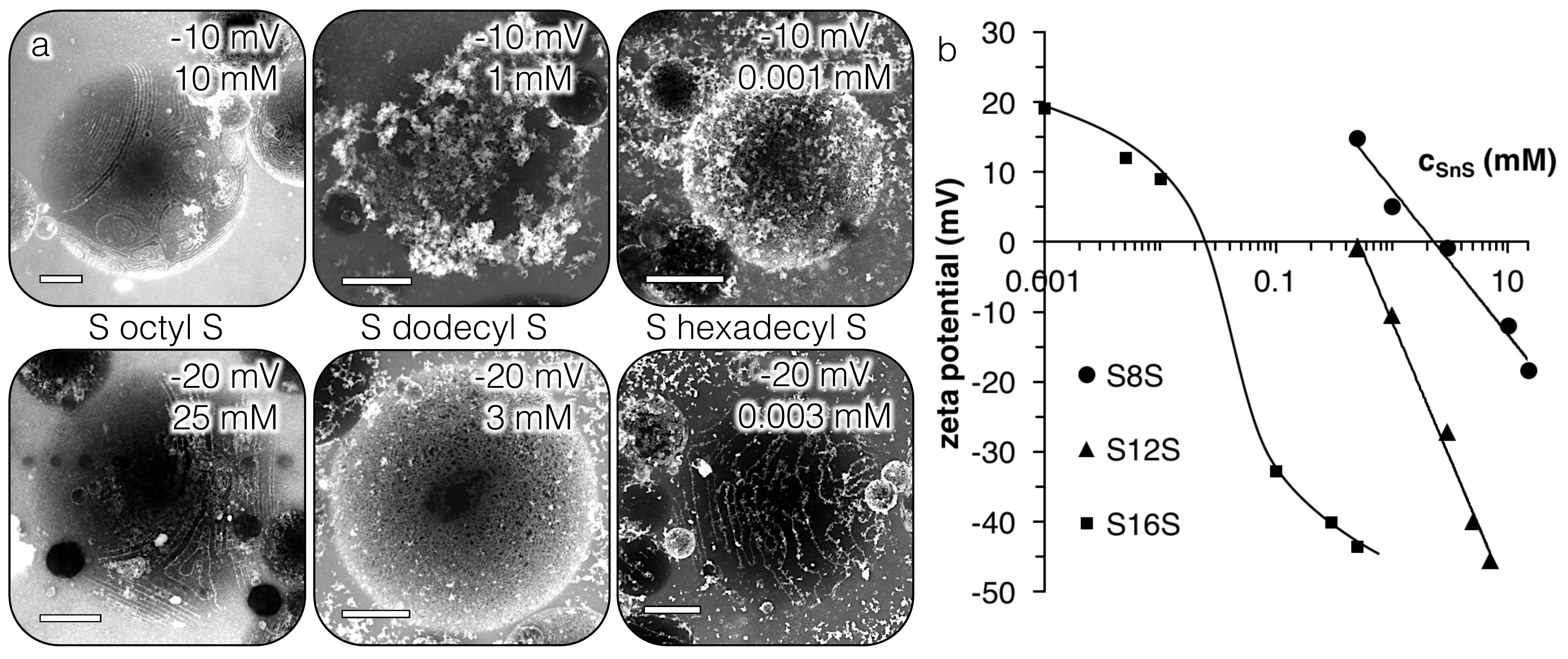}}
\caption{\label{S-VaryTail} Behavior of 0.0025 wt-\% of 200-nm, amine-functionalized silica nanoparticles, modified by varying the tail length of SA$_n$S, on cholesteric liquid crystal droplets in water (pH 7). Scale bars are 50 $\mu$m. Each column in a) represents differing tail lengths of sodium alkyl sulfate. For each row in a), the zeta potential of the nanoparticles are kept fixed to maintain approximately the same degree of SA$_n$S double-layer formation on the nanoparticle surfaces. For a surfactant tail length of $n=8$ (octyl), the nanoparticles are more dispersed than for $n=12$ (dodecyl) and 16 (hexadecyl), allowing them to better align with the homeotropic stripes of the cholesteric surface patterning. b) Zeta potential of 0.0025 wt-\% 200-nm, amine-functionalized silica nanoparticles with varying surfactant concentration for sodium octyl, dodecyl, and hexadecyl sulfate.} 
\end{figure*}

\begin{figure*}[ht!]
\centerline{\includegraphics[width=1\textwidth]{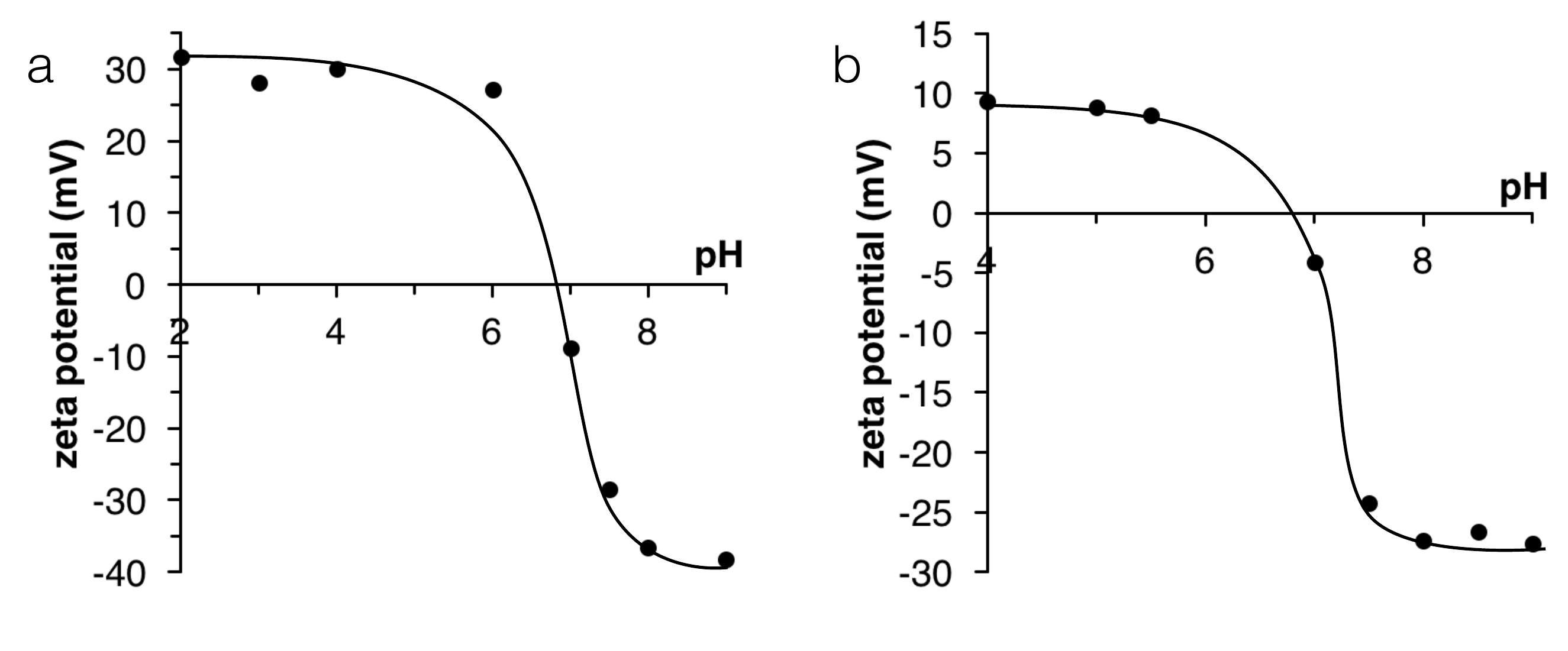}}
\caption{\label{S-Zeta} Zeta potential measurements of a) 200-nm amine-functionalized silica nanoparticles with 3 mM SOS and b) 30-nm silica nanoparticles with 10 mM C$_{8}$TAB, with varying pH from additions of either HCl or NaOH. Lines are drawn to guide the eye.} 
\end{figure*}

\begin{figure*}[ht!]
\centerline{\includegraphics[width=0.9\textwidth]{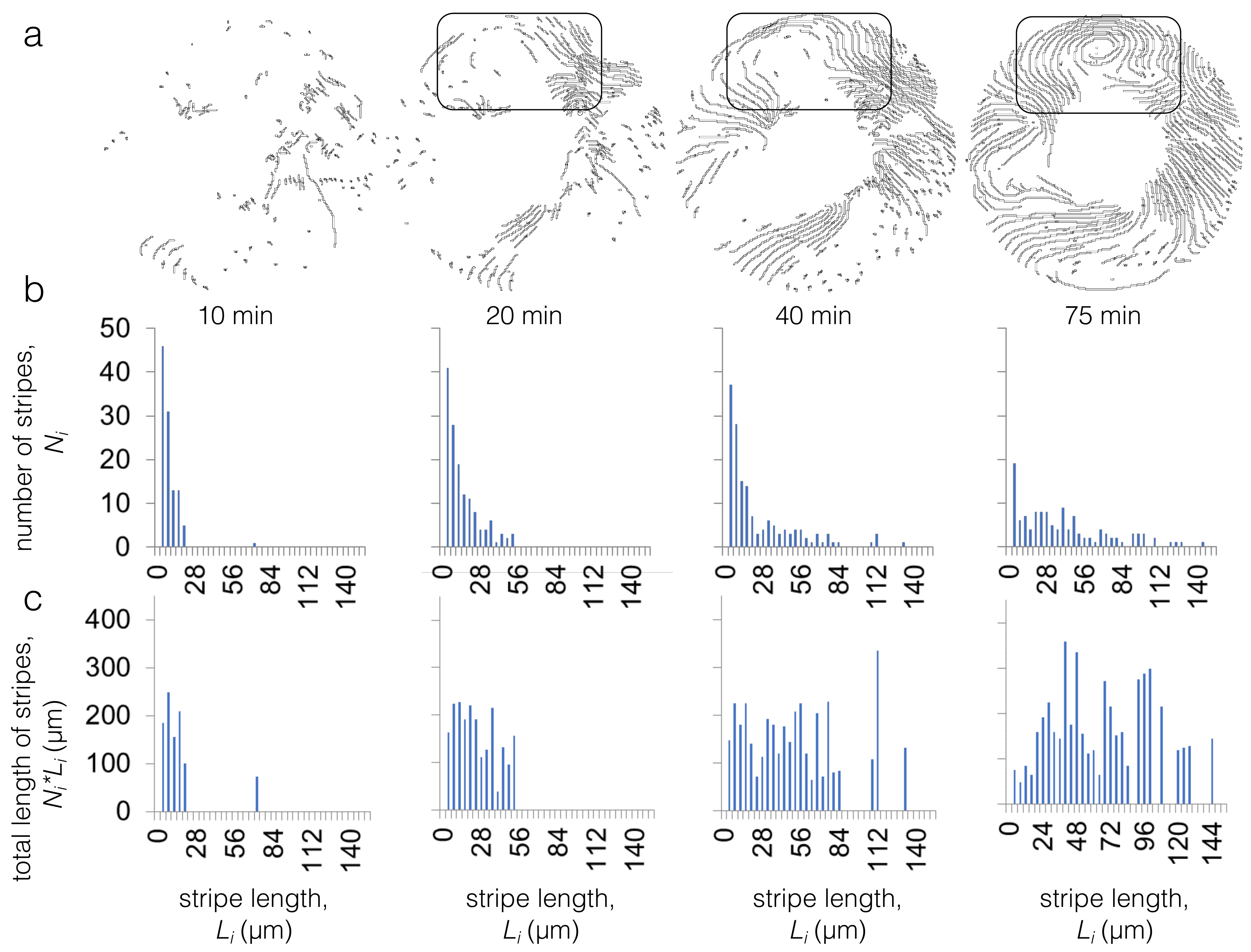}}
\caption{\label{S-TimeSeries} Analysis of the growth of nanoparticle-filled stripes on a cholesteric liquid crystal droplet. a) The nanoparticle assemblies are outlined and linearized through image analysis for each frame of the image sequence that captures their growth on a cholesteric droplet. The same droplet is described in Fig.~\ref{TimeEvo}a,b. b) The lengths of the nanoparticle stripes, $L_{i}$, are plotted against the frequencies of their occurrence, $N_{i}$, for each frame. The results after 10, 20, 40, and 75 minutes are shown. The frequency of smaller stripes is greater at the beginning of the interfacial nanoparticle attachment and decreases as the process progresses, suggesting a nucleation and growth mechanism. c) The total length of the stripes, $N_{i} \cdot L_{i}$, is also plotted against $L_{i}$ for each frame. These data are used to graph the total length and the number averaged length against time, plotted in Fig.~\ref{TimeEvo}c.} 
\end{figure*}

\begin{figure*}[ht!]
\centerline{\includegraphics[width=1\textwidth]{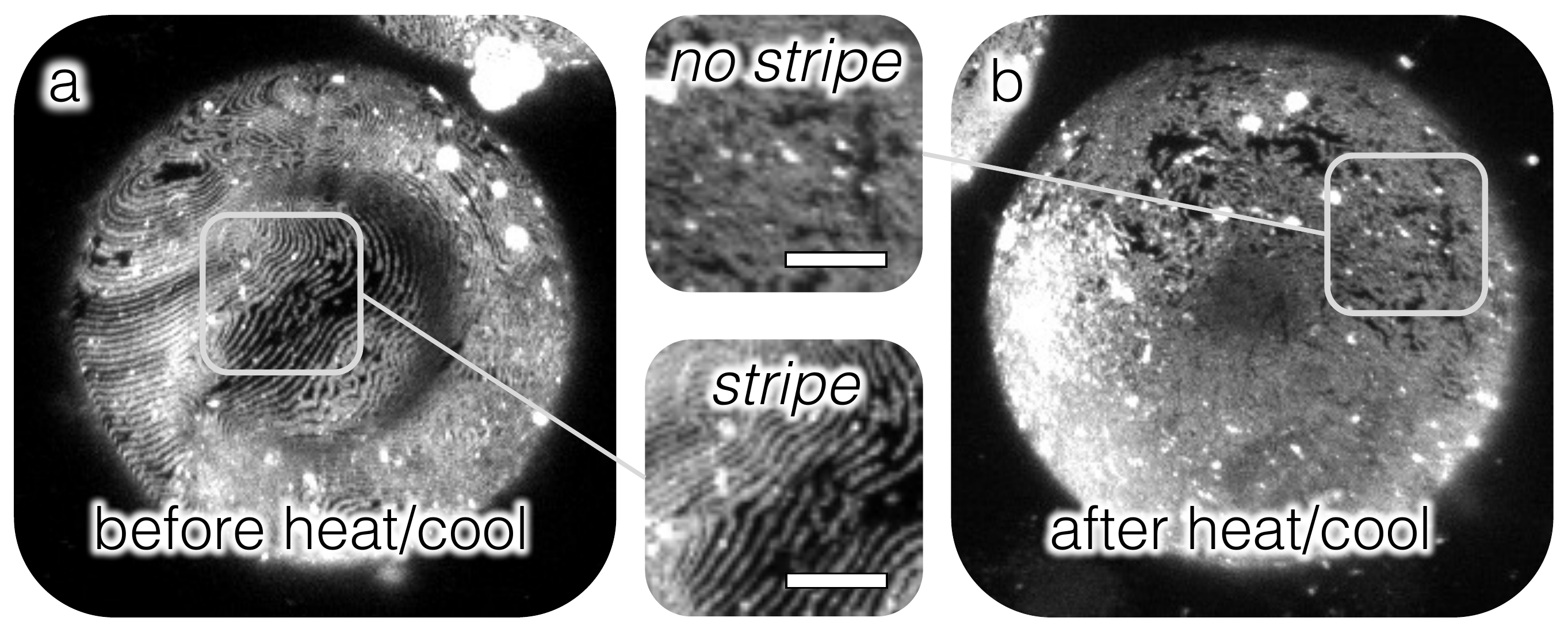}}
\caption{\label{S-HeatCool} Destroying cholesteric ordering with temperature. Scale bars are 25 $\mu$m. a) A confocal micrograph shows the nanoparticles on a cholesteric droplet without crosslinking. b) The confocal micrograph of the droplet in a) after quenching reveals that the nanoparticle assemblies are more disordered and generally do not retain their initial linear shape.} 
\end{figure*}

\end{document}